\documentclass[10pt,letterpaper,twocolumn]{article} %% two column, final layout

\usepackage{ol2}
\usepackage[draft]{hyperref}
\usepackage{amsmath}
\begin{document}

\twocolumn[ %% activate for two-column option

\title{Increasing the imaging capabilities of multimode fibers by exploiting the properties of highly scattering media }

\author{Ioannis N. Papadopoulos,$^{*}$ Salma Farahi, Christophe Moser and Demetri Psaltis}

\address{
School of Engineering, \'{E}cole Polytechnique F\'ed\'eral de Lausanne (EPFL), Station 17, 1015, Lausanne, Switzerland\\
$^*$Corresponding author: ioannis.papadopoulos@epfl.ch
}

\begin{abstract}We present a novel design that exploits the focusing properties of scattering media to increase the resolution and the working distance of multimode fiber based imaging devices. Placing a highly scattering medium in front of the distal tip of the multimode fiber enables  the formation of  smaller sized foci at increased working distances away from the fiber tip. We perform a parametric study of the effect of the working distance and the separation between the fiber and the scattering medium on the focus size. We experimentally demonstrate submicron focused spots as far away as 800$\mu$m with 532nm light. 
\end{abstract}

\ocis{060.2350, 070.5040, 090.1995, 110.7050.}

 ]

\noindent 
Endomicroscopy is a powerful technique that can help clinicians in diagnosis and surgery~\cite{flusberg2005fiber}. Among the different designs suggested for endoscopes, clinically available ones are generally based on fiber bundles were each single mode fiber core acts as a single pixel of the final image~\cite{Gmitro:1993gi}. Another category of endoscopes is built by combining a single mode fiber for the delivery of the excitation light, conventional optics and mechanical actuators for the focusing and scanning and finally a multimode fiber for the collection of the fluorescent signal~\cite{Gbel:2004dx}. Recently, considerable attention has been drawn towards multimode fibers that appear as promising candidates for high-resolution minimally invasive endoscopic imaging~\cite{Bianchi:2012du,Cizmar:2012gz,Choi:2012dx,Papadopoulos:2013uka}. The large number of degrees of freedom present in a multimode fiber allows for the transmission of large amounts of information, while at the same time its size remains relatively small. The overall diameter of the multimode fiber device can be limited to less than 500$\mu$m.

The new family of multimode fiber (MMF) based endoscopes proposed recently, can be used with different working distances. However, as was demonstrated in ~\cite{Cizmar:2012gz} and \cite{Papadopoulos:2013uka}, there is a clear tradeoff between working distance and the achievable resolution. This is because the resolution is directly linked to the Numerical Aperture (NA), therefore as the physical aperture of the MMF remains constant, the image quality achieved by such a system deteriorates as the working distance increases. 

It has been demonstrated that scattering media under the appropriate control of the incident wavefront can be used to focus light~\cite{Vellekoop:2010bp,Hsieh_digital_phase}, transmit an image~\cite{Popoff:2010dh},  rotate the polarization~\cite{Guan:12} or act as mirrors~\cite{Katz:2012ha}. Based on wavefront shaping techniques, even opaque layers of scattering media can be rendered transparent as far as the information transmission is concerned.  The optical field propagating through a scattering medium on its way to a focus spot on the other side of the scatterer, gets coupled to a broader spectrum of spatial frequencies compared to the incident beam. Van Putten et al.~\cite{vanPutten:2011jz} have exploited this effect to demonstrate super-resolved images of gold nanoparticles where the imaging device was substituted by a thin, highly scattering layer of GaAs, which under the appropriate wavefront control was transformed into the equivalent of a focusing optical lens.

In this paper, we demonstrate that a scattering medium used in conjunction with a multimode fiber imaging system can increase the working distance and the associated field of view while maintaining a high lateral resolution. Our approach is based on the synergetic exploitation of a multimode fiber and a highly scattering medium placed in front of the fiber, so that the effective number of degrees of freedom of the system is increased. We use Digital Phase Conjugation (DPC)~\cite{Bellanger:2008gs, Cui:2010wr, Hsieh_digital_phase,Papadopoulos:2012ko} to calculate the appropriate wavefront that will bring light into focus after it propagates through the fiber and the scattering medium.

\begin{figure}[htb]
\centerline{\includegraphics{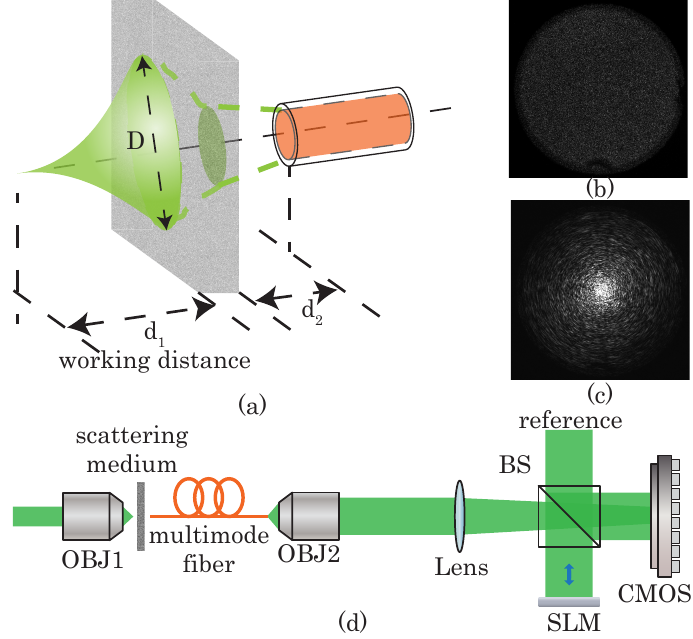}}
\caption{(a) A highly scattering medium placed in front of a multimode fiber will lead to a system with an increased number of degrees of freedom. The fiber diffracts with an angle associated with the fiber NA and enters the scattering medium where it gets diffused. The large aperture where the focusing originates from is a result of these two processes. (b) and (c). Comparison of the speckled output of the fiber with and without the scattering medium. The presence of the scattering medium causes the maximum number of modes to be excited in the fiber in a uniform manner (shown in (b)). (d) Experimental setup of Digital Phase Conjugation (DPC).}
\label{fig:1}
\end{figure}
The experimental implementation of the suggested scheme is depicted in Fig.~\ref{fig:1}. A highly scattering medium is placed in front of the distal tip of the multimode optical fiber at a specific distance d$_2$. The scattering medium used is a single layer of white paint of ~20$\mu$m thickness deposited on a glass slide (150$\mu$m thick). The working distance, d$_1$, of the composite system is defined as the focusing distance away from the end surface of the glass slide. The experimental implementation of DPC is the same as the one described in ~\cite{Papadopoulos:2012ko}. A 532nm continuous wave DPSS laser source was used to generate a diffraction-limited spot at the desired distance away from the scattering medium. The optical field propagates through the ensemble of the scattering medium and the multimode fiber and reaches the proximal tip of the fiber. The output of the fiber is imaged and holographically recorded on a CMOS digital sensor; the phase is extracted through the digital reconstruction in the computer and then assigned onto the phase--only SLM device. The reference beam is reflected off the SLM picking up the calculated phase information thus forming the optical phase conjugate beam. The phase conjugate beam back propagates first through the multimode fiber and then through the scattering medium forming a sharp focused spot at the original position. 

As can be seen in Fig. 1(a), the presence of the scattering medium has a double effect on the final aperture of the system. Initially the optical field exits the fiber and is diffracted with an angle equal to the NA of the fiber and as it propagates inside the turbid medium experiences multiple scattering and exits with an increased aperture compared to the entering one. The size of the focused spot will be defined by the size of the final output aperture of the system and the working distance.

\begin{figure}[htb]
\centerline{\includegraphics{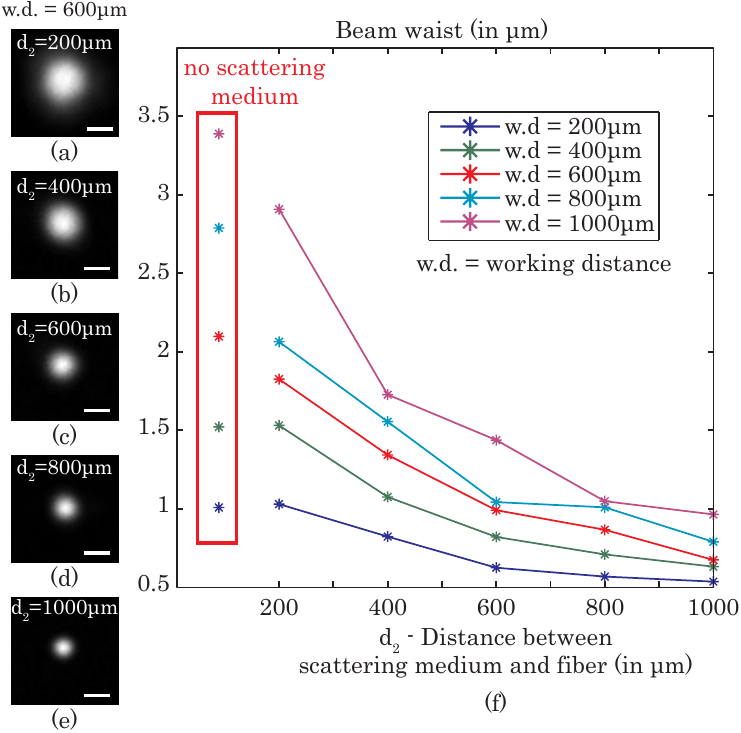}}
\caption{(a) through (e). Evolution of the beam size for a fixed working distance of 600$\mu$m as the distance between the scattering medium and the fiber increases from 200$\mu$m to 1000$\mu$m. The spot has the expected Gaussian profile and a beam waist smaller than 1$\mu$m is demonstrated. Scale bar is 1$\mu$m. (f) Quantitative results of the Gaussian fitted beam waist as a function of the distance between the fiber and the scattering medium for different working distances. As expected, the beam waist decreases when the fiber is moved away from the scattering.}
\label{fig:2}
\end{figure}

In Fig. 1(b) and (c) we compare the speckle output of the fiber at the proximal tip with and without the scattering medium. We can observe that the presence of the scattering medium causes a greater number of modes to be excited in the fiber in a uniform manner manifested by the uniform smaller speckles that are shown in Fig. 1(b).

Based on the above physical description, we can predict that for a fixed focal distance, the size of the spot will become smaller as the distance between the fiber end and the scatterer is increased. On the other hand, for a fixed distance between the fiber and the scattering medium, the aperture of the system remains constant and therefore the beam waist of the focussed spot will increase with increasing working distance.

In order to better quantify this behavior, we perform a parametric study of the focus size vs. the distances d$_1$ and d$_2$. The resulting focused spot after DPC is imaged onto a CCD sensor and the measured data are used as an input to a Gaussian curve-fitting algorithm. The calculated beam waist is defined as the 2*w$_0$ parameter of the fitted Gaussian profile.

The results demonstrated in Fig. 2, verify the expected behavior. Figures 2 (a) to (e) show the evolution of the generated focus at a working distance of 600$\mu$m as the distance between the scattering medium and the fiber is changed from 200$\mu$m to 1000$\mu$m. The focused spots have a shape very close to the ideally expected Gaussian. On the right hand side of Fig. 2, a summarized plot of all the measured results is presented. As it can be observed, for any certain working distance of the final device, an increase in the distance between the fiber and the scatterer signifies a decrease in the beam waist of the generated spot. The effect is very prominent especially when the focus is generated at 800$\mu$m and 1000$\mu$m, where we can see that spot sizes even smaller that 1$\mu$m are achieved. 

The effect of placing a scattering medium in front of the fiber output is to increase the effective NA of the system by directly increasing the physical output aperture where the focus is generated from. The effect in this sense, would be the same as placing a lens in front of the fiber. However, unlike a conventional lens, a free space scattering medium behaves in a spatially invariant stochastic way, meaning that the number of modes that will be excited in the scattering medium when any given optical field propagates through does not depend on the relative lateral position of the optical field to the scattering medium.

\begin{figure}[htb]
\centerline{\includegraphics{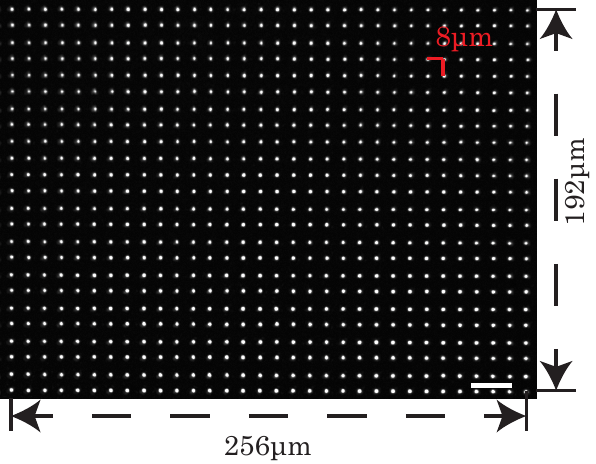}}
\caption{Superimposed images of focused spots generated across a 256$\mu$m x 192$\mu$m field of view, at a working distance of 600$\mu$m (distance between fiber and diffuser, d$_2$ = 600$\mu$m and distance between adjacent spots = 8$\mu$m). The size and quality of the spots remains constant across the whole field of view. The spatial invariance of the scattering medium leads to the excitation of the same number of modes and therefore the system is able to render aberration free foci across a wide field of view.  Scale bar, 20$\mu$m. }
\label{fig:3}
\end{figure}

In Fig. 3 a superimposed image of digitally scanned focused spots across a field of view of 256$\mu$m by 192$\mu$m is presented. We can observe that the quality and size of the generated foci does not degrade as the spot is moved away from the center and towards the edges of the field of view. As a result of this, the application of DPC on the combination of a scattering medium and a multimode fiber transforms it into a device capable of rendering high quality, aberration free, focused spots across a large field of view.

In order to perform imaging with the suggested technique, the major prerequisite is that the generated focused spot must have a sufficient SNR compared to the background. As can be verified by the images shown and measurements performed, the estimated SNR (defined as the maximum of the focus over the mean of the background) is in average higher than 2000:1. A potential difficulty arising from the suggested geometry is the fact that the scattering phenomena both during focusing and also during light collection for imaging decrease the expected photon budget. For the scattering medium used in the presented experiments the transmission attenuation was measured to be around 25dB. This can be overcome by increasing the power of the excitation laser source or by using a scatterer with a smaller scattering coefficient. A different approach could be to use a notch filter with a surrounding annular scattering ring in a dark-field configuration. This approach will exploit the annular ring for the generation of the focused spot while on the collection part, most of the fluorescent light will propagate backwards through the fiber. 

For practical applications, we propose the use of a deployable, “umbrella-like” mechanism in front of the multimode fiber, that will open up to the final form after the fiber endoscope is inserted inside the area that is to be imaged. In this configuration, the scattering medium bearing mechanism on the endoscope head can also prevent the deposition of tissue debris while the endoscopic head is inserted in the body. Also, we can imagine that a layer of tissue inside the area to be imaged can be used as the scatterer needed. In this case however, the presence of a beacon that will probe the scattering medium is necessary.

In conclusion, we suggest a configuration where a free space scattering medium and a multimode fiber work together to increase the resolution and working distance of fiber based imaging systems. We have demonstrated that this configuration can lead to the generation of submicron focused spots as far as 800$\mu$m away from the endoscopic head and can deliver aberration-free  tightly focused spots at an extended field of view of 256$\mu$m by 192$\mu$m. 

\vspace{6pt}
The authors acknowledge partial support from the Bertarelli Foundation under the grant, ``Optical Imaging of the inner ear for cellular diagnosis and therapy: Cochlear implants and beyond''.

\pagebreak

\end{document}